\documentclass[10pt,conference]{IEEEtran}
\IEEEoverridecommandlockouts
\usepackage{comment}
\usepackage{bm}
\usepackage{cite}
\usepackage{amsmath,amssymb,amsfonts}
\usepackage{algorithmic}
\usepackage{graphicx}
\usepackage{textcomp}
\usepackage{xcolor}
\usepackage{multirow}
\usepackage{color}
\usepackage{colortbl}
\usepackage{balance}
\usepackage{flushend}
\usepackage{multicol}
\usepackage[bookmarks=false]{hyperref}

\def\BibTeX{{\rm B\kern-.05em{\sc i\kern-.025em b}\kern-.08em
    T\kern-.1667em\lower.7ex\hbox{E}\kern-.125emX}}
\renewcommand{\thefootnote}{}

\usepackage{hyperref}
\makeatletter
\def\UrlAlphabet{%
      \do\a\do\b\do\c\do\d\do\e\do\f\do\g\do\h\do\i\do\j%
      \do\k\do\l\do\m\do\n\do\o\do\p\do\q\do\r\do\s\do\t%
      \do\u\do\v\do\w\do\x\do\y\do\z\do\A\do\B\do\C\do\D%
      \do\E\do\F\do\G\do\H\do\I\do\J\do\K\do\L\do\M\do\N%
      \do\O\do\P\do\Q\do\R\do\S\do\T\do\U\do\V\do\W\do\X%
      \do\Y\do\Z}
\def\UrlDigits{\do\1\do\2\do\3\do\4\do\5\do\6\do\7\do\8\do\9\do\0}
\g@addto@macro{\UrlBreaks}{\UrlOrds}
\g@addto@macro{\UrlBreaks}{\UrlAlphabet}
\g@addto@macro{\UrlBreaks}{\UrlDigits}
\makeatother

\begin{document}

\title{The Role of User Reviews in App Updates: A Preliminary Investigation on App Release Notes$^*$}
\author{\IEEEauthorblockN{Chong Wang$^\dagger$, Tianyang Liu$^\dagger$, Peng Liang}
\IEEEauthorblockA{School of Computer Science,
Wuhan University\\
Wuhan, China \\
\{cwang, leoii22, liangp\}@whu.edu.cn}

\and
\IEEEauthorblockN{Maya Daneva, Marten van Sinderen}
\IEEEauthorblockA{University of Twente \\
Enschede, the Netherlands \\
\{m.daneva, m.j.vansinderen\}@utwente.nl}
}
\maketitle

\renewcommand\footnoterule{\noindent \rule[0.25\baselineskip]{250pt}{0.5pt}}

\renewcommand{\thefootnote}{\fnsymbol{footnote}} 
\footnotetext[1]{This work is supported by the National Natural Science Foundation of China under Grant Nos. 61702378, 61972292, and 62032016.}
\footnotetext[2]{The authors contributed equally to this work. } 

\begin{abstract}

Release planning for mobile apps has recently become an area of active research. Prior research in this area concentrated on the analysis of release notes and on tracking user reviews to support app evolution with issue trackers. However, little is known about the impact of user reviews on the evolution of mobile apps. Our work explores the role of user reviews in app updates based on release notes. For this purpose, we collected user reviews and release notes of Spotify, the ‘number one’ app in the ‘Music’ category in Apple App Store, as the research data. Then, we manually removed non-informative parts of each release note, and manually determined the relevance of the app reviews with respect to the release notes. We did this by using Word2Vec calculation techniques based on the top 80 app release notes with the highest similarities. Our empirical results show that more than 60\% of the matched reviews are actually irrelevant to the corresponding release notes. When zooming in at these relevant user reviews, we found that around half of them were posted before the new release and referred to requests, suggestions, and complaints. Whereas, the other half of the relevant user reviews were posted after updating the apps and concentrated more on bug reports and praise.
\end{abstract}




\section{Introduction}

With the rapid progress on mobile techniques and smart phones, the number of mobile applications (apps for short) rises dramatically every year. As of the first quarter of 2021, Android users were able to choose between 3.48 million apps, making Google Play the app store with biggest number of available apps. The Apple App Store was the second-largest app store with roughly 2.22 million available apps for iOS~\cite{Ref1}. In these app repositories, users are free to post praises, complaints, and requests as reviews or ratings to the apps. Whereas, developers are responsible to provide apps with descriptions as well as app updates as release notes. This review-or-notes-posting activity makes app repositories now become the main data source to construct app datasets for the research on app development, evolution, and maintenance \cite{Ref39}.

As the most common and widely used app data, user reviews and ratings have been exploited to obtain user requirements or app features for software evolution and maintenance \cite{Ref37} \cite{Ref38} \cite{Ref36}. Although informative user reviews were reported to have valuable information (e.g., feature requests and bug reports) for developers, there is still no clarity on whether user reviews are actually taken into account by developers and implemented in the new release. Some researchers traced informative user reviews to source code changes to support evolution of successful apps \cite{Ref20} \cite{Ref21}. However, these authors focused their exploration on open source apps, rather than all the mobile apps. In fact, it turns out that very few studies explored the relation between app reviews and release notes in practice.

Release notes are textual documents delivered to the clients regularly, with the new release of a software product. These official texts are usually written by developers in a standardized way to present the main or important updates of the current version of software applications. Unlike source code changes, release notes are completely generated by developers, easier to collect, and not limited to open source software. Thus, release notes can be deemed good candidates to highlight those characteristics of updated apps for exploring the factors that influence app updates from app reviews.



To our best knowledge, very limited research examined the influence of user reviews on app updates from the perspective of developers, based on app release notes. This study makes a first step towards bridging this gap of knowledge and intends to explore the role that user reviews play in app updates, according to app release notes. Specifically, we would like to investigate whether developers consider user reviews when updating mobile apps. If yes, how many and which types of reviews contribute to or respond to the new releases? The preliminary findings will reveal the characteristics of the user feedback that got attention from developers. Plus, it can help both researchers and practitioners understand the relation between official release notes and personalized user feedback for app maintenance and evolution in practice. 




\section{Background and Related Work}
\label{rw}
In the past few years, many researchers (.e.g.~\cite{Ref42}) have contributed to the extraction and analysis of app review for software maintenance and evolution. However, prior research on app reviews mainly concentrated on the extraction of user requirements and/or app features, in order to get the key issues or topics for developers in updating apps. For example, both AR-Miner~\cite{Ref2} and Casper \cite{Ref3} were proposed to extract user requirements from app reviews. 
Unlike these authors, the present work did not attempt to extract user needs from app reviews, but to look into how user reviews support app updates, based on app release notes.

On the other hand, existing studies on release notes intend to explore the characteristics and development trends of app updates, by analyzing the posting time and content of release notes. 
For example,
McIlroy et al. explored the update frequency of 10,713 mobile apps across 30 mobile app categories~\cite{Ref12}. These authors indicate that 14\% of the apps are updated frequently, while 45\% of these frequently-updated apps do not provide users with any information about the rationale for the new updates. Plus, McIlroy et al. observed that frequently-updated apps are highly ranked by users. Compared to these studies, our concerns are not only the release notes and their attributes, but their relevance in regard to app reviews.

Other researchers tried to match user reviews with bug reports written by developers. For example, Häring et al.~\cite{Ref20} proposed an automatic approach DeepMatcher to match problem reports in app reviews to bug reports in issue trackers by using deep learning algorithms. Palomba et al. proposed CRISTAL~\cite{Ref21} to trace informative crowd reviews into code changes, in order to monitor the extent to which developers accommodate crowd requests and follow-up user reactions as reflected in their ratings. Their results indicate that developers implementing user reviews are rewarded in terms of ratings. However, the data sources of these two studies~\cite{Ref20}\cite{Ref21} have some limitation: the studies narrowed down their exploration to bug reports in issues trackers and source code changes from a few open-source projects. 
In addition, we found that the reports in issue trackers are expressed as a set of issues (e.g., bugs and feature requests) from various sources, and it is difficult to know whether these issues are addressed by developers. In contrast, we collected both release notes and user reviews from the same apps in our study. It is easy to identify which issues reported in user reviews got adopted or resolved in app updates, based on app release notes.

Villarroel et al.~\cite{Ref41} proposed an approach called CLAP (Crowd Listener for release Planning) to categorize user reviews, cluster together related reviews and prioritize the clusters of reviews to suggest app release planning. In a follow-up study, Scalabrino et al.~\cite{Ref40} improved CLAP for app developers to plan for the next release by selecting the most important raised complaints in a particular issue type. For evaluation, they used release notes as the basis, aiming to verify whether the issues clustered and prioritized by CLAP were resolved by the developers in subsequent updates. Unlike these authors' studies~\cite{Ref41}\cite{Ref40}, our work is based on the release notes, trying to identify the most relevant reviews to determine the degree of relevance and response between release notes and user reviews.

\section{Motivation and pilot study}
\label{motivation}
To investigate whether and how developers respond to user reviews based on app release notes, our work intends to explore how many release notes report the same issues as that mentioned in user reviews. For this purpose, we first calculated the similarity between a certain release note and the collected set of user reviews. Then, the user reviews are sorted in a descending order of similarity. Let N be a number that serves as a threshold for relevance in the ordered list of app reviews: for a chosen N, it will mean that the top N reviews are deemed to be relevant to this release note. Finally, we manually evaluated the relevance between each of the top N reviews and the specified release note in order to identify and remove the False Positive reviews, and to get the actual hit ratio of user reviews to a certain release note. However, we note that some parts of release notes at the sentence level are non-informative. To get a higher hit ratio of user reviews, we first conducted a pilot study to explore whether the hit ratio would be improved if the non-informative parts are removed from the release notes to be matched before calculating the similarity between release notes and user reviews. 

\begin{table*} [h]
	\caption{Top 5 relevant reviews of an exemplary release note with/without non-informative parts (R=Relevant, IR=IrRelevant)}
	\label{tab_pilot}
	\centering
	\begin{tabular}{|p{8.6cm}|p{0.5cm}|p{6.9cm}|p{0.5cm}|}
		\hline
		\textbf{Top 5 relevant user reviews for the raw release note} & \textbf{R/IR} & \textbf{Top 5 relevant user reviews for the processed release note} & \textbf{R/IR} \\
		\hline
        Navigation Menu was better off on the left side of the screen IPad & R &  Navigation Menu was better off on the left side of the screen IPad & R \\
        \hline
        Before the update I could pull up a menu by swiping up from the bottom on my iPhone screen to access my camera flashlight etc  & IR & Static menus at the bottom of the screen EFFICIENCY & R\\
        \hline
        I have an iPhone s and I don t have a very big screen and now that there are two bars at the bottom one for the tabs and one that has the song controls it makes my screen seem so small that I may as well be using an iPhone & IR & Love the menu on the bottom of the screen & R\\
        \hline
        The control panel is usually on the left side but then it & R & Bring back the static menus at the bottom of the screen & R \\
        \hline
        Static menus at the bottom of the screen EFFICIENCY & R & When I open the app the navigation menu on the bottom of the screen are missing & R \\
		\hline
	\end{tabular}
\end{table*}

To illustrate the process described above, we provide an example of the release note -- `\textit{ipad users you will find your navigation menu at the bottom of the screen just like iphone}'. The top five most relevant user reviews are listed in the first column of Table~\ref{tab_pilot}. By removing the non-informative parts, such as `just like iphone', `users', and stopwords, the processed release note has the other top five relevant reviews, as shown in the third column of Table~\ref{tab_pilot}. After manual evaluation, we found that the release note without non-informative parts would have a higher hit ratio of user reviews. R and IR in the second and fourth columns denote the results after manual evaluation. Table~\ref{tab_pilot} shows that after removing the non-informative parts, the hit ratio of relevant user reviews increases greatly. Because of this increase, later in Section \ref{result}, we will present results that are based only on the processed release notes by removing non-informative parts manually. 

\section{Research Design}

Our preliminary exploration intends to answer a single research question (RQ): \textit{\textbf{What are the roles that user reviews play in app updates, according to app release notes?}} Specifically, we want to know whether developers consider user reviews when updating mobile apps. If yes, how many and which types of reviews contribute to the new releases? Answering this RQ would help us understand the value of user reviews for app maintenance and evolution in practice.

\subsection{Data Collection}
\label{data_collection}
The dataset of this study is composed of app reviews and release notes of one mobile app, i.e., \textit{Spotify} in the `Music' category in Apple App Store \cite{Ref27}. The reason for choosing this app is traceable to the fact that \textit{Spotify} is the top free-download app in `Music' category in the region of US. As the most popular mobile application for streaming music, it has hundreds of thousands of reviews but not many pieces of release notes. This makes it well suited for our preliminary manual analysis. The data collection was performed in January 21, 2021, which returned 115 release notes and 450,381 user reviews between January 1, 2015 and January 21, 2021, since the first release note of \textit{Spotify} was posted on May 20, 2015.


\subsection{Pre-processing}
\label{data_processing}
To facilitate the automatic analysis in this study, the collected 450,381 app reviews and 115 release notes of \textit{Spotify} were pre-processed by means of the following steps. 


\textbf{Step 1: Cut app reviews and release notes into sentences.} 
To facilitate the analysis on app data at the sentence level, the collected 450,381 app reviews and 115 release notes were decomposed into 937,768 review sentences and 105 release note sentences, respectively, without duplicates.

\textbf{Step 2: Pre-process app reviews and release notes at sentence level.} In this step, multiple Natural Language Processing (NLP) techniques were applied to the textual contents of app reviews and release notes at sentence level. 
Specifically, Natural Language Toolkit (NLTK) was used to remove stopwords and punctuation and perform lemmatization on app reviews and release notes at sentence level.

\textbf{Step 3: Filter informative app reviews and release notes.} In this step, Logistic Regression (LR) and Linear Support Vector Machine (SVM) were applied to filter informative app reviews and release notes, respectively, according to their performance on 4,000 labelled app release notes in \cite{Ref5} and 4,000 labeled app reviews in \cite{Ref17}. This led to 50 informative release notes and 237,655 informative user reviews at the sentence level as the input of our analysis. 


\textbf{Step 4: Remove non-informative parts of release notes.} As mentioned in Section \ref{motivation}, removing non-informative parts of the release notes can significantly improve the hit ratio of user reviews that are relevant to specified release note.


\subsection{Matching user reviews to release notes}
\label{matching}
In this study, sentence similarity calculated by Word2Vec \cite{Ref18} is introduced to measure the degree of correlation between app reviews and release notes at the sentence level. 
The main reason for choosing this method \cite{Ref18} is that the other widely used similarity measures for text documents, such as Jaccard similarity coefficient, Edit Distance (Levenshtein Distance) \cite{Ref16}, and TF-IDF \cite{Ref33} coefficient, are all based on statistics, and cannot support semantic similarity matching in this work.

In particular, this study performs the calculation of sentence similarity in the most regular and common manner. That is, we first get the vector representation of each word in a sentence from Word2Vec. Then, the average of all the word vectors of this sentence is generated to get a representation of this sentence, by using Formula (1). Finally, the cosine similarity algorithm, as defined in Formula (2), is used to calculate the similarity between two sentences.

\begin{align}
    sentenceVector = \dfrac{\underset{i = wordNum}{\Sigma}wordVector_i}{wordNum}
\end{align}
where \textit{wordNum} is the number of words in a sentence and \textit{wordVector} is the word vector value of a word.

\begin{align}
    similarity = \dfrac{sentenceVector_i\cdot sentenceVector_j}{|sentenceVector_i|\cdot|sentenceVector_j|}
\end{align}
where \textit{sentenceVector} is the calculation result of the average word vector for each sentence.

Note that in this study, the calculation of sentence similarity was employed without any extension or improvement. However, the calculation of sentence similarity can be adapted to meet various purposes. Therefore, this study could also help identify the weakness of Word2Vec in the calculation of sentence similarity as well as the possible ways to improve similarity calculation between sentences.


\subsection{Manual selection on the most relevant reviews}
\label{top80}
Since the number of user reviews is much greater than the number of release notes, it is not feasible to review each of the user reviews and evaluate its relevance to the release notes manually. Inspired by the work of Haering et al.~\cite{Ref20}, we calculated the similarity between each user review and each of the 35 pre-processed release notes, and then took the top N reviews for manual labeling of relevance. In this study, we assigned N to be equal to 80. This means, we selected the top 80 matched reviews of each release note for manual validation of relevance.

\subsection{Manual labeling of relevance}
\label{RD_rq1}
Manual labeling for relevance on the top 80 reviews of each of the release notes was conducted by two coders independently, both of which are bachelor students major in software engineering. To make sure the two coders did the labeling in a consistent way, they compared their results. Whenever they had different labels, they discussed why there was a difference. The discussion was used to consolidate their understanding and arrive at a consensus on the label to be put. More specifically, for any given review, the label \textit{relevant} indicates that the content of this review reflects the release note, including users' praise, criticism and feedback on the update, users' requests, suggestions or complaints adopted by the developers, as well as users' reports on the bugs fixed by the developers. Whereas, \textit{irrelevant} means that the user review is mismatched.


\subsection{Types of relevant reviews}
\label{RD_rq2}
Furthermore, we intended to investigate what types of relevant user reviews either catch developers' eyes that contribute to app updates, or respond to app updates, according to app release notes. In this study, we defined six types of relevant reviews, i.e., \textit{praise}, \textit{dispraise}, \textit{bug report}, \textit{request and suggestion},\textit{ complaint}, as well as \textit{others}. These six types are adapted from 
\cite{Ref6} and their detailed descriptions are provided in Table~\ref{tab_label}. The manual labeling was conducted by the same two coders in a way similar to what is described in Section~\ref{RD_rq1}.


\begin{table} [htbp]
	\caption{Description of six types of relevant user reviews.}
	\label{tab_label}
	\centering
	\begin{tabular}{|p{0.1cm}|p{1.3cm}|p{5.8cm}|}
		\hline
		\textbf{\#} & \textbf{Label}& \textbf{Description}\\
		\hline
		1 & praise & expresses appreciation\\
		\hline
		2 & dispraise & opposite of praise\\
		\hline
		3 & bug report & bug report or crash report\\
		\hline
		4 & request and suggestion & asks for certain things, e.g. feature or improvement, and suggests acquisition\\
		\hline
		5 & complaint & feels discontent about certain things, e.g. missing feature or annoying bugs\\
		\hline
		6 & others & none of above\\
		\hline
	\end{tabular}
\end{table}

\section{Results and Discussion}
\label{result}
\subsection{Experimental data}
By following the four steps of data pre-processing in Section~\ref{data_processing}, 15 out of the 50 release notes could not find matched user reviews through Word2Vec, since all the words in those release notes, such as \textit{fix}, \textit{stability}, \textit{security} and \textit{issue}, are too general. Therefore, the final experimental dataset consists of 35 release notes and the corresponding 2800 user reviews (i.e., top 80 reviews per release note) that are automatically generated by Word2Vec.

\subsection{Roles of user reviews}
\begin{figure}[htbp]
	\includegraphics[width=\linewidth]{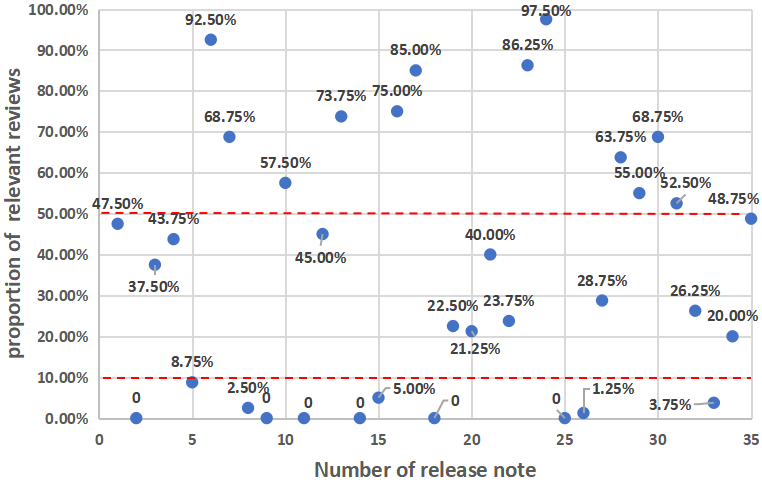}
	\caption{Distribution of the 35 release notes with their hit ratio.}
	\label{fig_rq1}
\end{figure}

By implementing the manual labeling for relevance defined in Section~\ref{matching}, we found that
among these 2,800 highly similar user reviews, a total of 1,042 were labeled as \textit{relevant}, and the remaining 1,758 were labeled as \textit{irrelevant}. For an overall, the hit ratio is 37.21\%. Figure~\ref{fig_rq1} summarizes the distribution of the 35 release notes and their corresponding hit ratio\footnote{Details are available at: \url{https://github.com/Leolty/Shared_Data/blob/main/35_release_notes.xlsx}}. We observed that 6 of the 35 release notes (i.e., R2, R9, R11, R14, R18, and R25) actually have no relevant user reviews and their hit ratios are 0\%. Whereas,
12 of the remaining 29 release notes are with a higher hit ratio (greater than 50\%) for matching relevant reviews. In addition, the hit ratio of the 35 release notes was distributed on average in the three intervals (divided with red dotted lines in Figure~\ref{fig_rq1}): [0,10\%), [10\%,50\%) and [50\%,100\%], each accounting for around one-third of the 35 release notes. 

Furthermore, considering the posting time of both the 1,042 relevant reviews and their corresponding release notes, we found that the posting time of 550 user reviews were earlier than that of the corresponding release notes. Whereas, 492 reviews were posted after the corresponding release notes. Therefore, we identified two roles of user reviews in app updates.\textit{ Pre-released reviews} (52.78\% of the 1,024 release notes) contributed to app updates, because they got developers' attention and even got implemented in the new releases. \textit{Post-released reviews} (47.22\%) serve as the feedback to the new releases and encourage the latest updating of this app. 

\textbf{Discussion:} There are around 37.2\% of the 2,800 matched user reviews (1042 reviews) actually relevant to the corresponding release notes. This indicates that although the most matched user reviews mentioned similar issues or topics to the release notes, only a small percentage of matched reviews actually got developers' attention and contributed to app updates. Plus, these 1,042 relevant reviews did not actually respond to all the 35 release notes. This makes us think that it is worthwhile to further investigate how to improve the matching rate between app reviews and release notes. In addition, more than a half of the relevant user reviews were posted before the corresponding new release of the app. This indicates that developers did take user reviews into account to some extent when updating the mobile apps. 

\subsection{Types of relevant user reviews}

\begin{figure}[htbp]
	\includegraphics[width=\linewidth]{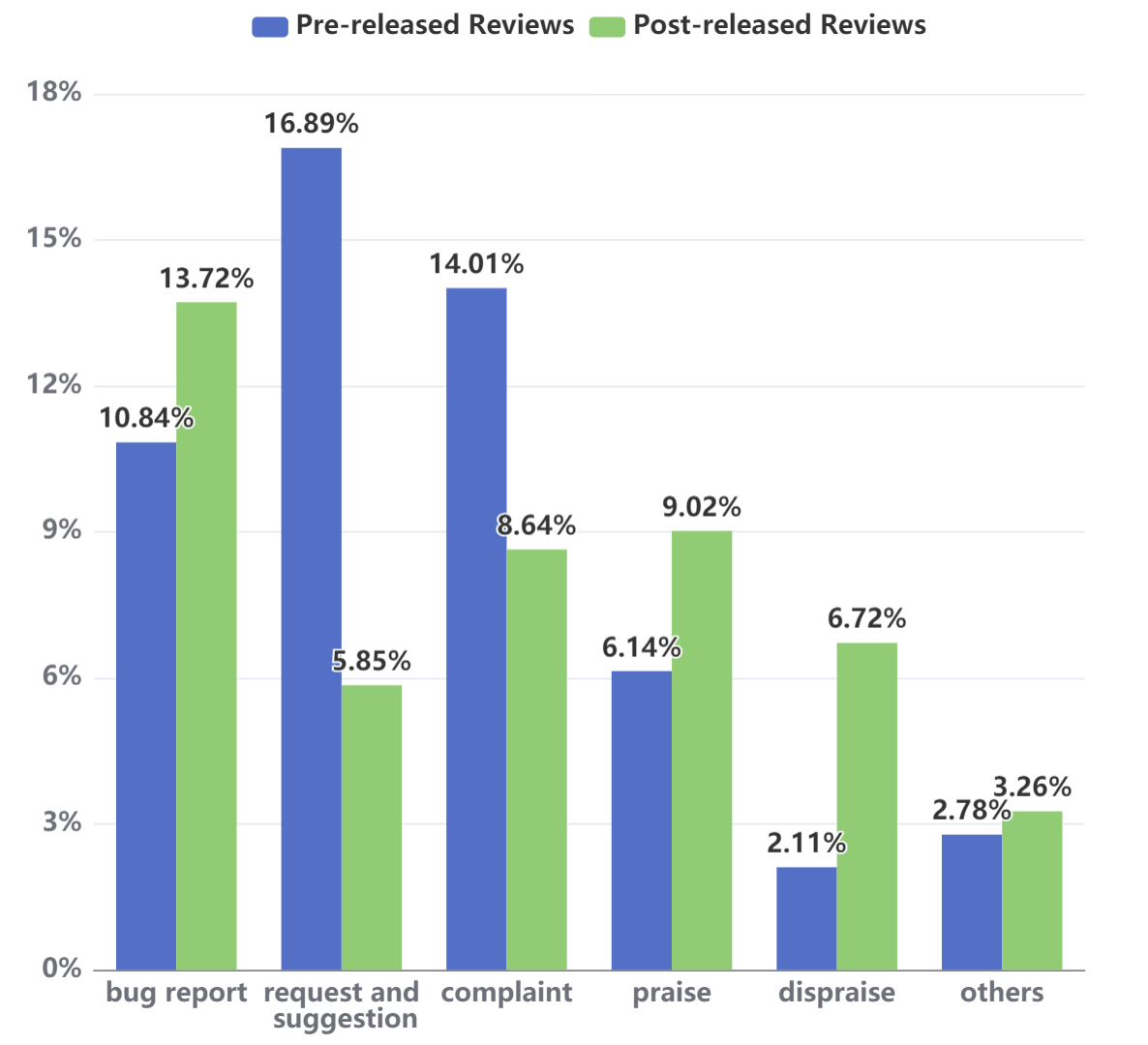}
	\caption{The percentage of 1,042 relevant user reviews over the six types and two roles (pre-released and post-released reviews).}
	\label{fig_rq2}
\end{figure}

Figure~\ref{fig_rq2} shows the percentages of 1,042 relevant user reviews over the six types defined in Section~\ref{RD_rq2} and the two roles (pre-released and post-released reviews). We found that \textit{bug report}, \textit{request and suggestion}, and \textit{complaint} are the top 3 most popular review types, covering 256, 237, and 236 reviews, respectively. Then, 158 user reviews refer to \textit{praise}, and 92 refer to \textit{dispraise}. \textit{others} reviews have relatively moderate proportion, covering 63 reviews. 

From the perspectives of two roles of user reviews in app updates, we found that in the 550 pre-released reviews, \textit{request and suggestion}, \textit{complaint}, and \textit{bug report} are the three most popular types, covering 176, 146, and 113 reviews, respectively. Only a few user reviews mentioned \textit{dispraise} and \textit{others}, accounting for 22 and 29 reviews, respectively. Regarding the 492 post-released reviews, \textit{bug report} (143 reviews), \textit{praise} (94 reviews) and \textit{complaint} (90 reviews) dominated over the other types. 

Moreover, as shown in Figure~\ref{fig_rq2}, it is obvious that pre-released reviews focus more on \textit{request and suggestion} and \textit{complaint}, while the post-release reviews are mainly with \textit{praise} and \textit{dispraise}. Regarding the remaining two types, i.e., \textit{bug report} and \textit{others}, the percentages have very little difference in pre-released and post-released user reviews.

\textbf{Discussion:}
Regarding the distribution of 1,042 relevant reviews over the six types, we found that \textit{bug report}, \textit{request and suggestion}, and \textit{complaint} are the three most mentioned types in the app reviews responded by developers.
Moreover, we found that the focuses of pre-released and post-released reviews differ. This is also consistent with our understanding that users often express their needs as complaints or requests and suggestions in pre-released user reviews. However, it is also realistic to expect that the introduction of new features may bring new bugs. This, in turn, is likely to lead to the largest number of bug reports in the post-released reviews. Plus, as users' feedback on new releases, \textit{dispraise} and \textit{praise} mainly occur in the post-release reviews.



\section{Limitations}
This study has some limitations. First, regarding the generalizability \cite{Ref10} of our findings, this work only included one mobile app, i.e., \textit{Spotify}. Although our results are not generalizable to all apps available in the market, we might possibly observe similar findings in the apps similar to Spotify. The similarities of these apps, for example, could be with similar services and functionalities, a similar user base with a similar culture, attitude and expectations on quality of other Music apps.  

Second, the release notes of \textit{Spotify} are concise. The proposed methods for data pre-processing and analysis may not be applicable for those apps with overly simple or overly detailed release notes.

Finally, although Word2Vec outperformed most of the other similarity measures in this study, its performance was not good enough as expected because 22 out of the 50 release notes are reported to have no actual relevant reviews.
Particularly, when applying Word2Vec to calculate the similarity between user reviews and release notes, we observed a high jargon barrier between these two app data sources. That is, some key words or jargon in app release notes, such as \textit{timestamp} or \textit{VoiceOver}, are not in the corpus of the Word2Vec model. This results in failed matching of relevant reviews to release notes. Therefore, more similarity measures or improvements on existing measures are highly desirable and are expected to provide solutions for polysemy and sequentiality in textual documents of mobile apps.
In addition, in this work, we averaged the vectors of all the words in a sentence to calculate the cosine similarity between two sentences. This may lead to match few similar app review sentences, if there are lots of non-informative words in the sentences to be matched. In order to achieve a high matching accuracy, 
weighted average of the word vectors can be introduced by assigning a higher weight to the same or similar words and/or phrases (including synonyms and antonyms) in two sentences, when calculating the similarity between them. 

\section{Conclusions and Future Work}
This study sheds light on the role that user reviews play in updating mobile apps, based on app release notes. We found that (1) for the investigated app, i.e., \textit{Spotify}, user reviews are indeed taken into account by the developers when updating the apps. However, only 37.2\% of the hit user reviews are actually relevant to the corresponding release notes; and (2) 52.78\% of these relevant user reviews were posted before the corresponding release notes and focused on requests, suggestions and complaints, whereas 47.22\% are post-released reviews, paying more attention to bug reports and praise. 

The next steps of this preliminary study are as follows: (1) to expand the size of the experimental dataset, (2) to improve the similarity calculation method according to our research need, (3) to minimize the jargon barrier between official release notes and personalized user reviews, and (4) to give guidance on how to write valuable user reviews.

\balance
\bibliographystyle{ieeetr}
\bibliography{main}

\begin{thebibliography}{10}

\bibitem{Ref1}
Statista, ``Number of apps available in leading app stores as of 3rd quarter
  2020.''
  \url{https://www.statista.com/statistics/276623/number-of-apps-available-in-leading-app-stores/},
  2020.

\bibitem{Ref39}
W.~Martin, F.~Sarro, Y.~Jia, Y.~Zhang, and M.~Harman, ``A survey of app store
  analysis for software engineering,'' {\em Proceedings of IEEE Transactions on
  Software Engineering}, vol.~43, no.~9, pp.~817--847, 2017.

\bibitem{Ref37}
F.~Palomba, P.~Salza, A.~Ciurumelea, S.~Panichella, H.~Gall, F.~Ferrucci, and
  A.~De~Lucia, ``Recommending and localizing change requests for mobile apps
  based on user reviews,'' in {\em Proceedings of the 39th IEEE/ACM
  International Conference on Software Engineering (ICSE)}, pp.~106--117, 2017.

\bibitem{Ref38}
A.~Di~Sorbo, S.~Panichella, C.~V. Alexandru, C.~A. Visaggio, and G.~Canfora,
  ``Surf: Summarizer of user reviews feedback,'' in {\em Proceedings of the
  39th IEEE/ACM International Conference on Software Engineering Companion
  (ICSE-C)}, pp.~55--58, 2017.

\bibitem{Ref36}
S.~Panichella, A.~Di~Sorbo, E.~Guzman, C.~A. Visaggio, G.~Canfora, and H.~C.
  Gall, ``How can i improve my app? classifying user reviews for software
  maintenance and evolution,'' in {\em Proceedings of the 31st IEEE
  International Conference on Software Maintenance \& Evolution (ICSME)},
  pp.~281--290, 2015.

\bibitem{Ref20}
M.~Haering, C.~Stanik, and W.~Maalej, ``Automatically matching bug reports with
  related app reviews,'' in {\em Proceedings of the 43rd IEEE/ACM International
  Conference on Software Engineering (ICSE)}, pp.~970--981, 2021.

\bibitem{Ref21}
F.~Palomba, M.~Linares-Vásquez, G.~Bavota, R.~Oliveto, and A.~D. Lucia, ``User
  reviews matter! tracking crowdsourced reviews to support evolution of
  successful apps,'' in {\em Proceedings of the 31st IEEE International
  Conference on Software Maintenance \& Evolution (ICSME)}, pp.~291--300, 2015.

\bibitem{Ref42}
N.~Genc-Nayebi and A.~Abran, ``A systematic literature review: Opinion mining
  studies from mobile app store user reviews,'' {\em Journal of Systems \&
  Software}, vol.~125, pp.~207--219, 2017.

\bibitem{Ref2}
N.~Chen, J.~Lin, S.~C.~H. Hoi, X.~Xiao, and B.~Zhang, ``Ar-miner: Mining
  informative reviews for developers from mobile app mark app marketplace,'' in
  {\em Proceedings of the 36th ACM International Conference on Software
  Engineering (ICSE)}, pp.~767--778, 2014.

\bibitem{Ref3}
H.~Guo and M.~P. Singh, ``Caspar: Extracting and synthesizing user stories of
  problems from app reviews,'' in {\em Proceedings of the 42nd ACM
  International Conference on Software Engineering (ICSE)}, pp.~628--640, 2020.

\bibitem{Ref12}
S.~Mcilroy, N.~Ali, and A.~E. Hassan, ``Fresh apps: an empirical study of
  frequently-updated mobile apps in the google play store,'' {\em Empirical
  Software Engineering}, vol.~21, no.~3, pp.~1346--1370, 2016.

\bibitem{Ref41}
L.~Villarroel, G.~Bavota, B.~Russo, R.~Oliveto, and M.~Penta, ``Release
  planning of mobile apps based on user reviews,'' in {\em Proceedings of the
  38th International Conference on Software Engineering (ICSE)}, pp.~14--24,
  2016.

\bibitem{Ref40}
S.~Scalabrino, G.~Bavota, B.~Russo, M.~D. Penta, and R.~Oliveto, ``Listening to
  the crowd for the release planning of mobile apps,'' {\em IEEE Transactions
  on Software Engineering}, vol.~45, no.~1, pp.~68--86, 2019.

\bibitem{Ref27}
Apple, ``Apple app store.'' \url{https://apps.apple.com/us/genre/ios/id36mt=8}.

\bibitem{Ref5}
C.~Wang, J.~Li, P.~Liang, M.~Daneva, and M.~V. Sinderen, ``Developers' eyes on
  the changes of apps: An exploratory study on app changelogs,'' in {\em
  Proceedings of the 3rd International Workshop on Crowd-Based Requirements
  Engineering (CrowdRE)}, pp.~207--212, 2019.

\bibitem{Ref17}
M.~Lu and P.~Liang, ``Automatic classification of non-functional requirements
  from augmented app user reviews,'' in {\em Proceedings of the 21st ACM
  International Conference on Evaluation and Assessment in Software Engineering
  (EASE)}, pp.~344--353, 2017.

\bibitem{Ref18}
T.~{Mikolov}, K.~{Chen}, G.~{Corrado}, and J.~{Dean}, ``{Efficient Estimation
  of Word Representations in Vector Space},'' {\em arXiv e-prints},
  p.~arXiv:1301.3781, Jan. 2013.

\bibitem{Ref16}
V.~I. Levenshtein, ``Binary codes capable of correcting deletions, insertions,
  and reversals,'' in {\em Soviet Physics Doklady}, vol.~10, pp.~707--710,
  Soviet Union, 1966.

\bibitem{Ref33}
L.-P. Jing, H.-K. Huang, and H.-B. Shi, ``Improved feature selection approach
  tfidf in text mining,'' in {\em Proceedings of the 1st IEEE The International
  Conference on Machine Learning and Cybernetics (ICMLC)}, vol.~2,
  pp.~944--946, 2002.

\bibitem{Ref6}
D.~Pagano and W.~Maalej, ``User feedback in the appstore: An empirical study,''
  in {\em Proceedings of the 21st IEEE International Conference on Requirements
  Engineering (RE)}, pp.~125--134, 2013.

\bibitem{Ref10}
R.~Wieringa and M.~Daneva, ``Six strategies for generalizing software
  engineering theories,'' {\em Science of Computer Programming}, vol.~101,
  no.~apr.1, pp.~136--152, 2015.

\end{thebibliography}


\end{document}